\documentclass[english,aps,preprint,superscriptaddress]{revtex4-1}
\usepackage{lmodern}
\usepackage{lmodern}
\usepackage[T1]{fontenc}
\usepackage{amsmath}
\usepackage{amssymb}
\usepackage{graphicx}
\usepackage{esint}

\makeatletter
\usepackage{amsthm}\usepackage{latexsym}\usepackage{bm}\usepackage{amsfonts}
\usepackage{babel}

\usepackage{babel}

\usepackage{hyperref}

\makeatother

\usepackage{babel}
\begin{document}

\title{Structure-preserving geometric particle-in-cell algorithm suppresses
finite-grid instability -- Comment on ``Finite grid instability
and spectral fidelity of the electrostatic Particle-In-Cell algorithm''
by Huang et al.}

\author{Jianyuan Xiao}

\affiliation{School of Physical Sciences, University of Science and Technology
of China, Hefei, 230026, China}

\author{Hong Qin}
\email{hongqin@princeton.edu}

\affiliation{Plasma Physics Laboratory, Princeton University, Princeton, NJ 08543,
U.S.A}

\affiliation{School of Physical Sciences, University of Science and Technology
of China, Hefei, 230026, China}
\begin{abstract}
A recent paper by Huang et al. {[}Computer Physics Communications
207, 123 (2016){]} thoroughly analyzed the Finite Grid Instability
(FGI) and spectral fidelity of standard Particle-In-Cell (PIC) methods.
Numerical experiments were carried out to demonstrate the FGIs for
two PIC methods, the energy-conserving algorithm and the momentum-conserving
algorithm. The paper also suggested that similar numerical experiments
should be performed to test the newly developed Structure-Preserving
Geometric (SPG)-PIC algorithm. In this comment, we supply the results
of the suggested numerical experiments, which show that the SPG-PIC
algorithm is able to suppress the FGI.
\end{abstract}



\maketitle
\global\long\def\EXP{\times10^}  
\global\long\def\rmd{\mathrm{d}}  
\global\long\def\rmc{\mathrm{c}}  
\global\long\def\diag{\textrm{diag}}  
\global\long\def\xs{ \mathbf{x}_{sp}}  
\global\long\def\bfx{\mathbf{x}}  
\global\long\def\bfd{\mathbf{d}}  
\global\long\def\bfp{\mathbf{p}}  
\global\long\def\bfv{\mathbf{v}}  
\global\long\def\bfA{\mathbf{A}}  
\global\long\def\bfY{\mathbf{Y}}  
\global\long\def\bfB{\mathbf{B}}  
\global\long\def\bfS{\mathbf{S}}  
\global\long\def\bfG{\mathbf{G}}  
\global\long\def\bfE{\mathbf{E}}  
\global\long\def\bfM{\mathbf{M}}  
\global\long\def\bfQ{\mathbf{Q}}  
\global\long\def\bfu{\mathbf{u}}  
\global\long\def\bfe{\mathbf{e}}  
\global\long\def\bfzig{\mathbf{r}_{\textrm{zig2}}}  
\global\long\def\bfxzig{\mathbf{r}_{\textrm{xzig}}}  
\global\long\def\bfzzig{\mathbf{r}_{\textrm{zzig}}}  
\global\long\def\xzig{\mathbf{x}_{\textrm{zig}}}  
\global\long\def\yzig{\mathbf{y}_{\textrm{zig}}}  
\global\long\def\zzig{\mathbf{z}_{\textrm{zig}}}  
\global\long\def\zigspmvar{\left( \bfx_{sp,l-1},\bfx_{sp,l},\tau \right)}  
\global\long\def\zigspvar{\left( \bfx_{sp,l},\bfx_{sp,l+1},\tau \right)}  
\global\long\def\rme{\mathrm{e}}  
\global\long\def\rmi{\mathrm{i}}  
\global\long\def\rmq{\mathrm{q}}  
\global\long\def\ope{\omega_{pe}}  
\global\long\def\oce{\omega_{ce}}  
\global\long\def\FIG#1{Fig.~\ref{#1}}  
\global\long\def\TAB#1{Tab.~\ref{#1}}  
\global\long\def\EQ#1{Eq.~(\ref{#1})}  
\global\long\def\SEC#1{Sec.~\ref{#1}}  
\global\long\def\APP#1{Appendix~\ref{#1}}  
\global\long\def\REF#1{Ref.~\cite{#1}}  
\global\long\def\DDELTAT#1{\textrm{Dt}\left(#1\right)}  
\global\long\def\DDELTATA#1{\textrm{Dt}^*\left(#1\right)}  
\global\long\def\GRADD{ {\nabla_{\mathrm{d}}}}  
\global\long\def\CURLD{ {\mathrm{curl_{d}}}}  
\global\long\def\DIVD{ {\mathrm{div_{d}}}}  
\global\long\def\CURLDP{ {\mathrm{curl_{d}}^{*}}}  
\global\long\def\DIVDP{ {\mathrm{div_{d}}^{*}}}  
\global\long\def\cpt{\captionsetup{justification=raggedright }}  
\global\long\def\act{\mathcal{A}}  
\global\long\def\calL{\mathcal{L}}  
\global\long\def\calJ{\mathcal{J}}  
\global\long\def\DELTAA{\left( \bfA_{J,l}-\bfA_{J,l}' \right)}  
\global\long\def\DELTAAL{\left( \bfA_{J,l-1}-\bfA_{J,l-1}' \right)}  
\global\long\def\ADAGGER{\bfA_{J,l}^\dagger}  
\global\long\def\ADAGGERA#1{\bfA_{J,#1}^{x/2}}  
\global\long\def\EDAGGER#1{\bfE_{J,#1}^{x/2}}  
\global\long\def\BDAGGER#1{\bfB_{J,#1}^{x/2}}  
\global\long\def\DDT{\frac{\partial}{\partial t}}  
\global\long\def\DBYDT{\frac{\rmd}{\rmd t}}  
\global\long\def\DBYANY#1{\frac{\partial }{\partial #1}} 
\newcommand{\WZERO}[1]{W_{\sigma_0 I}\left( #1 \right)} 
\newcommand{\WONE}[1]{W_{\sigma_1 J}\left( #1 \right)} 
\newcommand{\WONEJp}[1]{W_{\sigma_1 J'}\left( #1 \right)} 
\newcommand{\WTWO}[1]{W_{\sigma_2 K}\left( #1 \right)} 
\newcommand{\WTHREE}[1]{W_{\sigma_3 L}\left( #1 \right)} 
\newcommand{\bfJ}{\mathbf{J}} 
\global\long\def\MQQ{M_{00}} 
\global\long\def\MDQDQ{M_{11}} 
\global\long\def\MDQQ{M_{01}} 

Huang et al. recently provided an in-depth analysis of the Finite
Grid Instability (FGI) and spectral fidelity of standard Particle-In-Cell
(PIC) methods \cite{huang2016finite}. The spectral errors, especially
the aliased spatial modes, from charge deposition and field interpolation
schemes were rigorously quantified. Numerical experiments were carefully
designed and carried out to demonstrate the FGIs of the Momentum-Conserving
(MC)-PIC algorithm \cite{birdsall1991plasma} and the Energy-Conserving
(EC)-PIC algorithm \cite{lewis1970energy}. Simulation results using
a Particle-And-Spectrum (PAS) method \cite{evstatiev2013variational}
was also given for comparison and benchmark. The paper suggested performing
similar numerical experiments to test the newly developed Structure-Preserving
Geometric (SPG)-PIC algorithm \cite{squire2012geometric,xiao2013variational,xiao2015explicit,he2015hamiltonian,he2016hamiltonian,xiao2016explicit,qin2016canonical,xiao2017local,kraus2017gempic,xiao2018structure}.
In this comment, we supply the results of the suggested numerical
experiments using the specific implementation of the SPC-PIC algorithm
reported in Ref.\,\cite{xiao2015explicit}.

The parameters for the numerical experiments are the same as in Ref.\,\cite{huang2016finite},
which are listed as follows. The simulation domain is a $L\times1\times1$
periodic box where $L=33$, $\Delta x=1$, and $\omega_{p}=2\pi/L$.
The time step is set to $\Delta t=0.2\omega_{p}^{-1}$. The numbers
of sampling points per grid for both electron and ions are $300$,
and the mass ratio and charge ratio between electrons and ions are
$1:3672$ and $-1:1$, respectively. Initially the ions are equally
spaced and their velocities are set to $0$. Electrons are equally
spaced with a sinusoidal displacement $\delta x\left(x_{p_{0}}\right)=LA\cos\left(2\pi Mx_{p_{0}}/L\right)/(2\pi M)$,
and their velocities are $v_{x}(x)=0.01\rmc+LA\omega_{p}\sin\left(2\pi Mx/L\right)/(2\pi M)$,
where $A=0.01$ and $M=9$. Initial electric field is $E_{x}\left(x\right)=-Lq_{e}A\cos\left(2\pi Mx/L\right)/\left(2\pi M\right)$.

\begin{figure}
\begin{centering}
\includegraphics[width=0.49\linewidth]{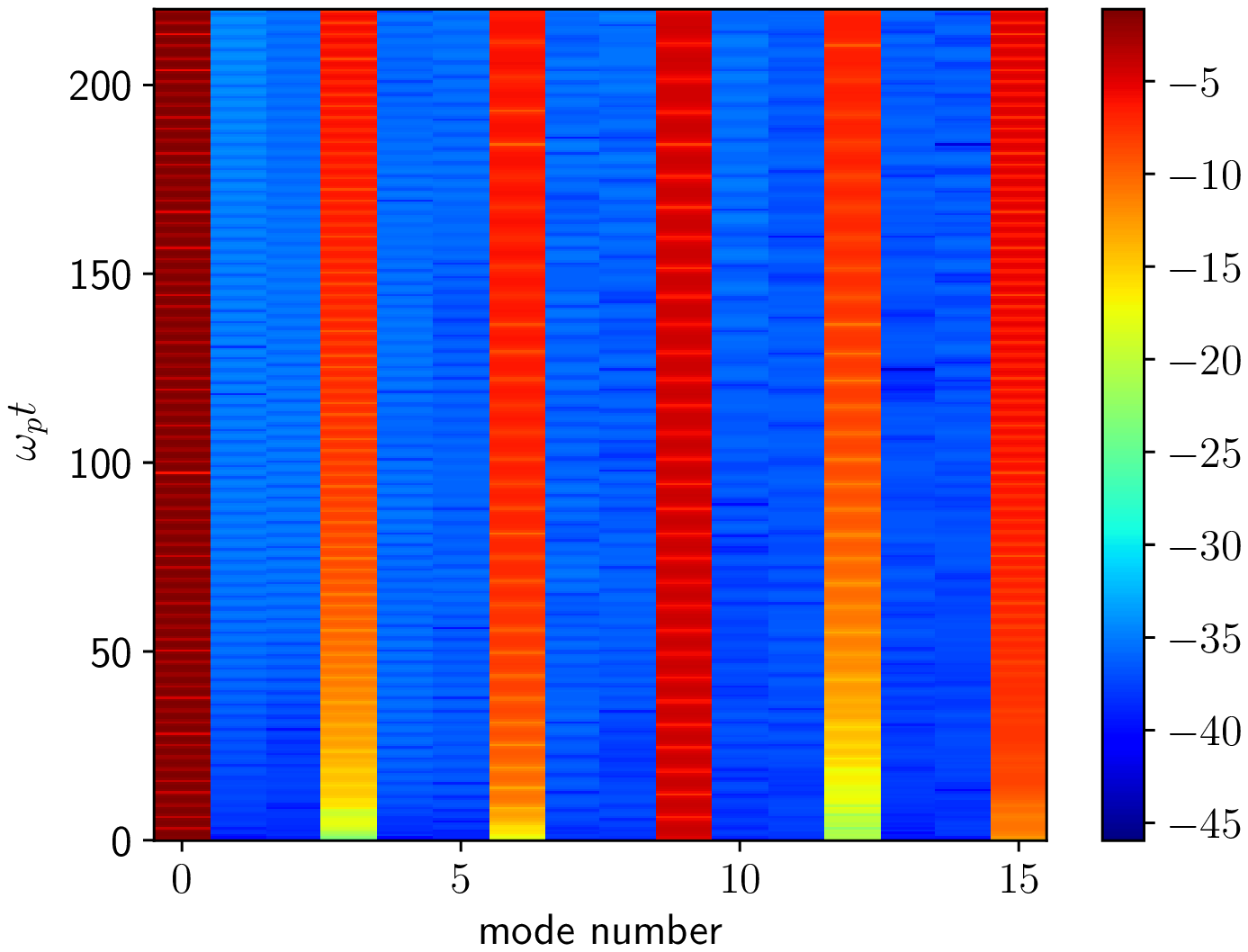}\includegraphics[width=0.49\linewidth]{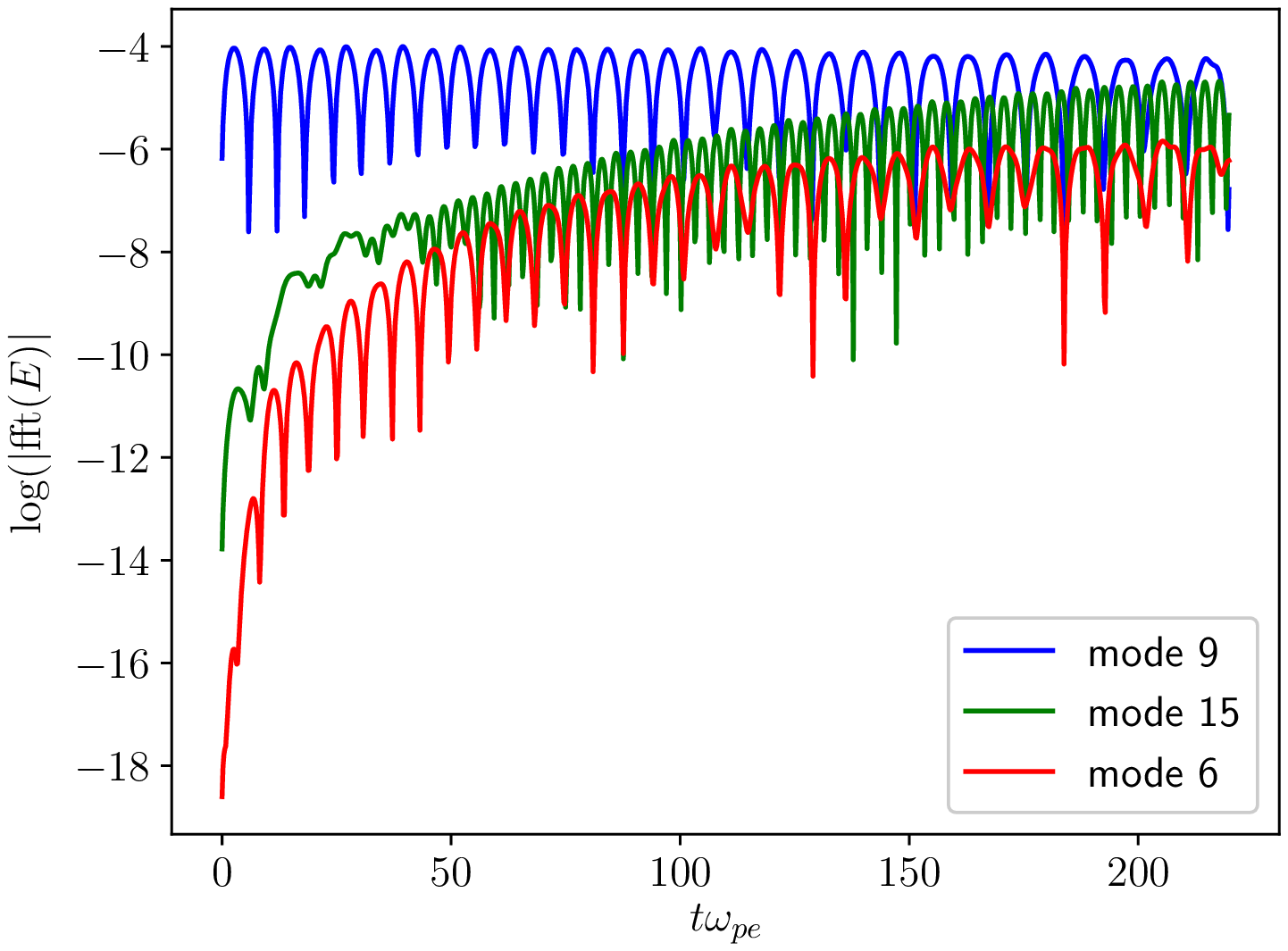}
\par\end{centering}
\caption{Mode spectrum simulated by the SPG-PIC algorithm. Full mode spectrum
$\log(|\mathrm{fft}(E)|)$ as a function of time is shown in the left
figure, and evolution of the mode amplitude for mode number $k=9,15,6$
is shown in the right figure.}
\label{FigEHSmod} 
\end{figure}

\begin{figure}
\begin{centering}
\includegraphics[width=0.49\linewidth]{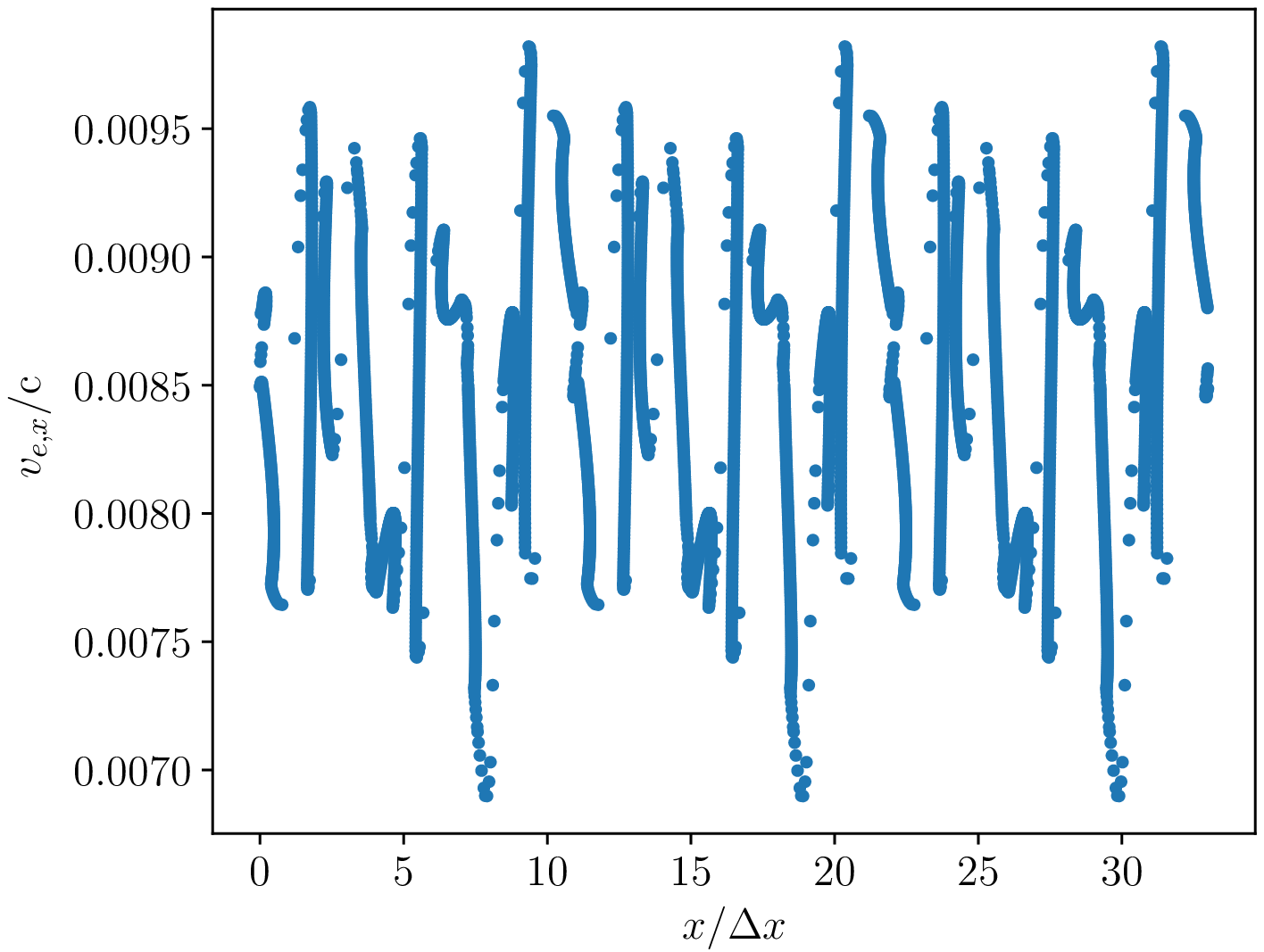}\includegraphics[width=0.49\linewidth]{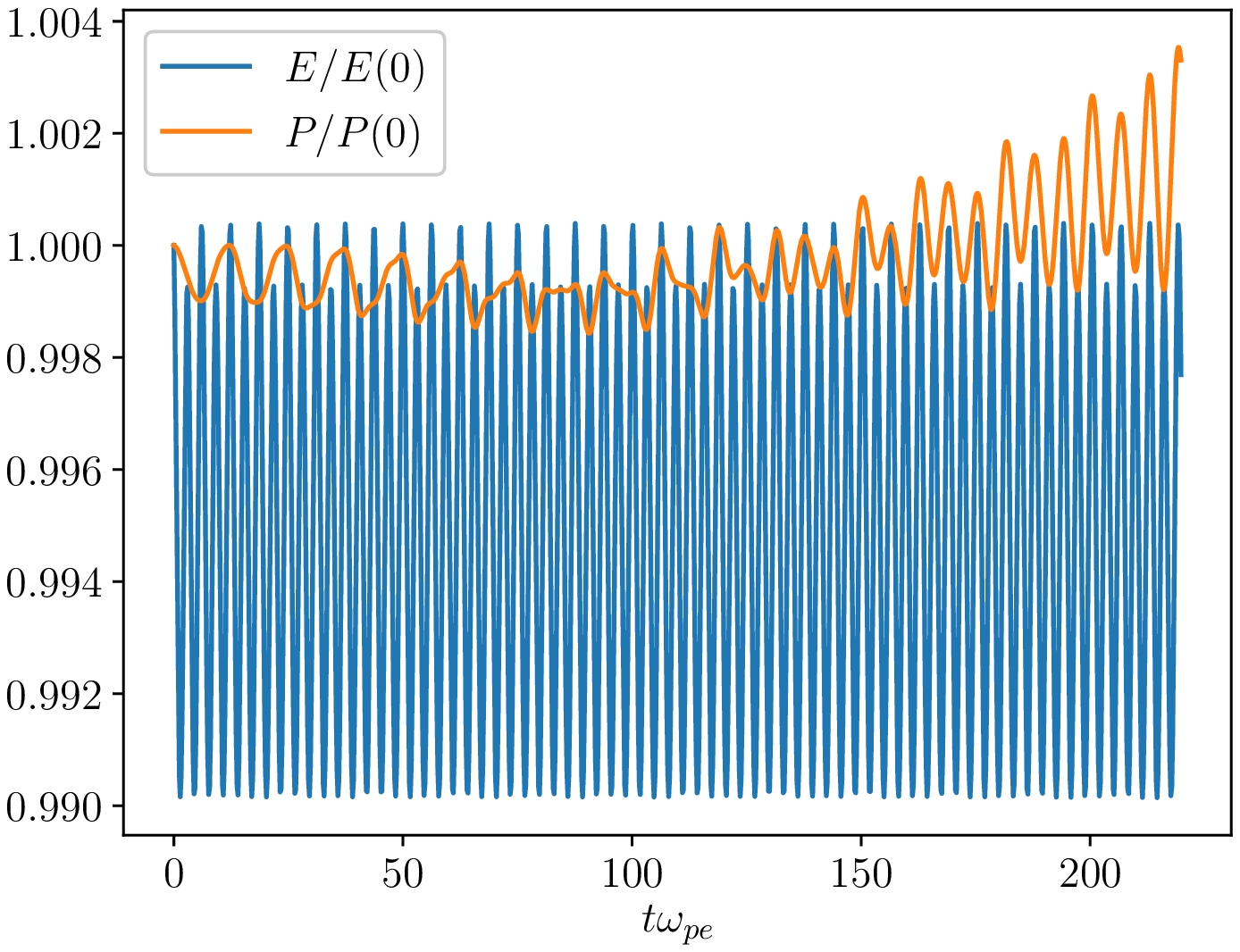}
\par\end{centering}
\caption{Electron distribution at $t=200\omega_{p}^{-1}$ (left) and the evolution
of energy and momentum (right). }
\label{FigEHSEXV} 
\end{figure}

The simulation first is performed to $t=220\omega_{p}^{-1}$. The
resulting mode spectrum, final velocity distribution, energy and momentum
evolution are plotted in Figs.\,\ref{FigEHSmod} and \ref{FigEHSEXV},
which correspond to Fig.\,3 and Fig.\,4 of Ref.\,\cite{huang2016finite},
respectively. These results show that even though mode alias still
exists in the SPG-PIC algorithm, the FGI is suppressed. While mode
alias effect, as an error of spatial discretization, is inevitable
in any spatial grid, its existence does not necessarily imply unstable
numerical eigenmodes will be exited. Note that when a unstable numerical
eigenmode is excited, all components of the dynamics, especially the
dominant ones, of the discrete system will grow exponentially. Unfortunately,
such FGIs do exist in the standard PIC methods. As demonstrated in
Ref.\,\cite{huang2016finite}, in about $30$ plasma oscillation
periods ($200\omega_{p}^{-1}$), the total energy error for the MC-PIC
algorithm exceeds $200\%,$ and the total momentum error for the EC-PIC
algorithm exceeds 70\%. On the other hand, Fig.\,\ref{FigEHSEXV}
shows that for the SPG-PIC algorithm, the total energy error is less
than $1\%,$ and the total momentum error is less than $0.4\%.$

The observed suppressing of FGI for the SPG-PIC algorithm can be attributed
to the structure-preserving nature of its spatial discretization.
The charge deposition and field interpolation are derived from a variational
principle using the techniques of Whitney interpolation forms \cite{squire2012geometric,xiao2016explicit,xiao2018structure}
or finite element discrete exterior calculus \cite{he2016hamiltonian,kraus2017gempic},
which preserves the discrete gauge symmetry and the discrete exterior
calculus structure of the electromagnetic field. As a result, physical
laws, such as the charge conservation and $\nabla\cdot\mathbf{B}=0$,
are satisfied exactly by the discrete system. This result is consistent
with Ref.\,\cite{huang2016finite}'s conclusion that charge deposition
and field interpolation can be optimally designed to suppress or reduce
FGIs.

\begin{figure}
\begin{centering}
\includegraphics[width=0.49\linewidth]{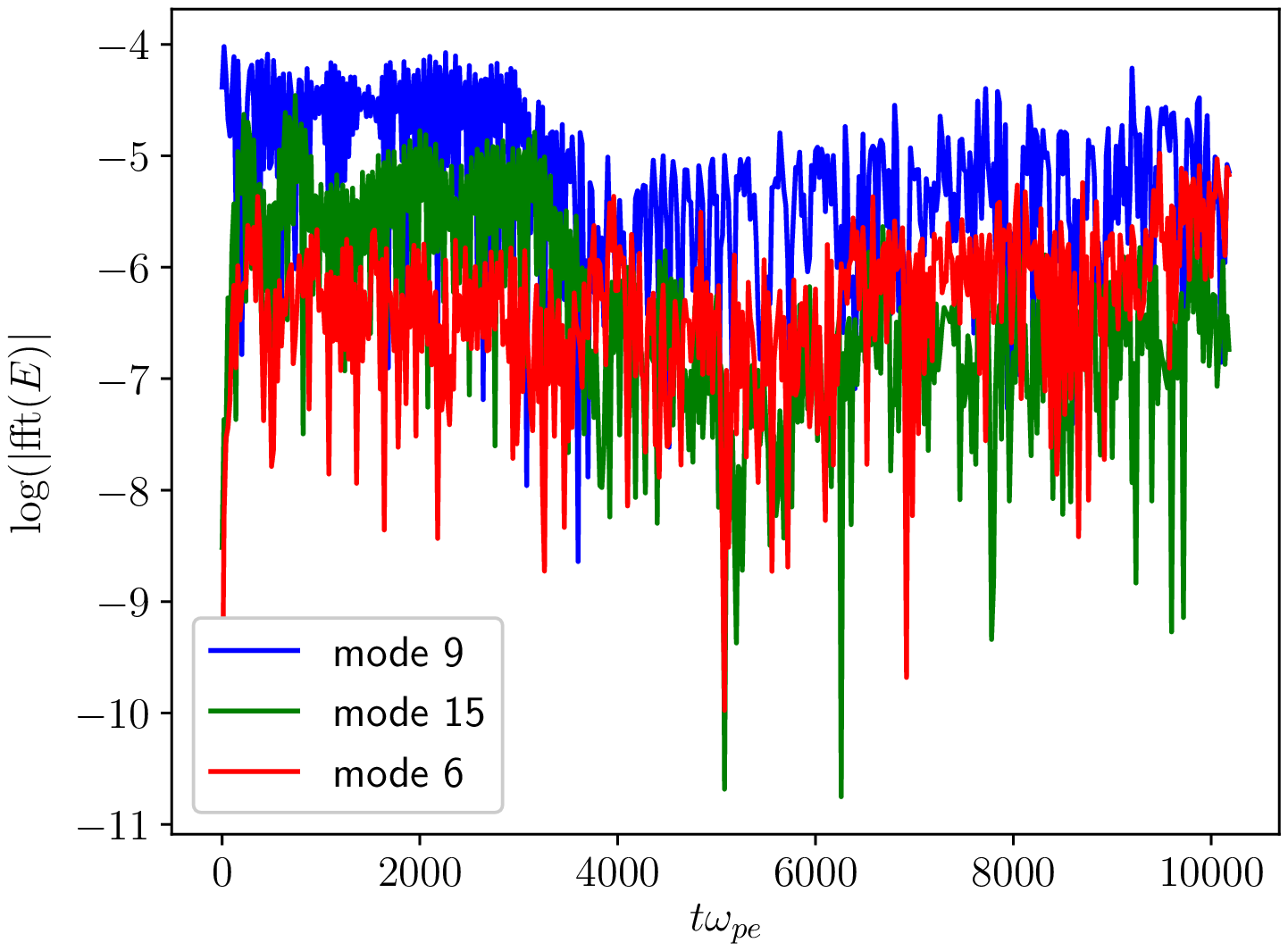}\includegraphics[width=0.49\linewidth]{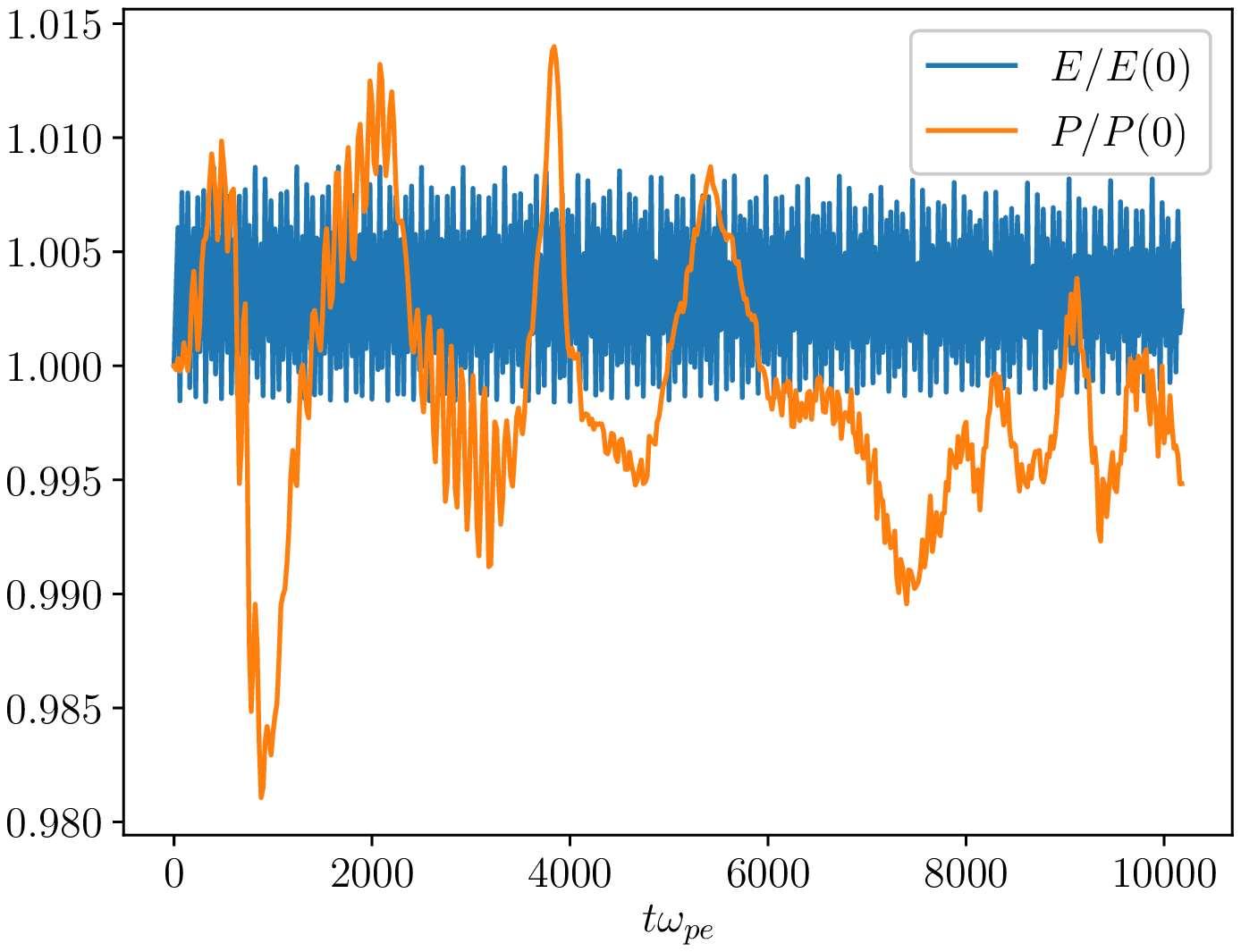} 
\par\end{centering}
\caption{Long term evolution of spectrum (left), total energy and momentum
(right) simulated by the SPG-PIC algorithm.}
\label{FigEHSLONG} 
\end{figure}

Another feature of the SPG-PIC algorithm is the preserving of non-canonical
symplectic structure for time-integration, which in general bounds
simulation errors on conserved quantities for a very long time. We
run the simulation for $500$ longer to $t=10000\omega_{p}^{-1}$,
and the result is plotted in Fig.\,\ref{FigEHSLONG}. Over this long
simulation time, the total energy error and total momentum error are
bounded by $1\%$ and $2\%$, respectively.

We finish this comment with two footnotes. First, the SPG-PIC algorithm
used is for Vlasov-Maxwell system in 3D configuration space. For the
simplified geometry and simulation parameters of the present numerical
experiments, the dominated modes of the discrete system are longitudinal
electrostatic modes. Secondly, for the PAS methods, symplectic time-integration
can also be adopted. For example, Cary and Doxas \cite{Cary93,Doxas1997}
first applied a canonical symplectic algorithm to simulate the particle-and-mode
Hamiltonian models \cite{Mynick1978,Escande1982,Escande87,Escande89,Escande91,Escande2018}
for the Vlasov-Poisson system.

\section*{Acknowledgments}

We thank Prof.\,John Cary and Prof.\,Dominique Escande for fruitful
discussions. This research is supported by the National Key Research
and Development Program (2016YFA0400600, 2016YFA0400601, 2016YFA0400602
and 2018YFE0304100), the National Natural Science Foundation of China
(NSFC-11775219, NSFC-11575186 and NSFC-11805273), China Postdoctoral
Science Foundation (2017LH002), the Fundamental Research Funds for
the Central Universities (WK2030040096) and the U.S. Department of
Energy (DE-AC02-09CH11466).

\bibliographystyle{apsrev4-1}
\bibliography{fgispic}

\end{document}